\documentclass[graybox]{svmult}

\usepackage{amsmath}
\usepackage{amssymb}

\usepackage{mathptmx}       
\usepackage{helvet}         
\usepackage{courier}        
\usepackage{type1cm}        
\usepackage{makeidx}         
\usepackage{graphicx}        
\usepackage{multicol}        
\usepackage[bottom]{footmisc}

\usepackage{color}
\usepackage{url}

\def\res{{\text{res}}}

\def\cD{{\mathcal{D}}}
\def\cW{{\mathcal{W}}}

\newcommand{\bN} { {\mathbb{N}}}

\newcommand{\bZ} { {\mathbb{Z}}}

\newcommand{\si} { {\sigma}}

\newcommand{\Tr}{{\operatorname{Tr}}}

\makeindex             

\begin{document}

\title*{Bivariate Extensions of Abramov's Algorithm for Rational Summation\\
\bigskip
{\small \text{Dedicated to Professor Sergei A. Abramov on the occasion of his 70th birthday}}}
\titlerunning{Bivariate Extensions of Abramov's Algorithm for Rational Summation}

\author{Shaoshi Chen}

\institute{Shaoshi Chen \at KLMM, AMSS, Chinese Academy of Sciences,  100190, Beijing,  China\\ \email{schen@amss.ac.cn} \smallskip \\
This work was supported by the NSFC grants 11501552, 11688101 and
by the President Fund of the Academy of Mathematics and Systems Science, CAS (2014-cjrwlzx-chshsh).}
%
%
\maketitle


\abstract{ Abramov's algorithm enables us to decide whether a univariate rational function can be written as
a difference of another rational function, which has been a fundamental algorithm for rational summation. In 2014,
Chen and Singer generalized Abramov's algorithm to the case of rational functions in two ($q$-)discrete variables.
In this paper we solve the remaining three mixed cases, which completes
our recent project on bivariate extensions of Abramov's algorithm for rational summation.
}

\section{Introduction} \label{SECT:intro}
Symbolic summation has been a powerful tool in combinatorics and mathematical physics, whose history
is as long as that of symbolic computation. Abramov's algorithm~\cite{Abramov1971} for rational summation is one of the first few fundamental
algorithms in symbolic summation. The central problem in symbolic summation is whether the sum of
a given sequence can be written in \lq\lq closed form\rq\rq. A given sequence~$f(n)$ belonging to some domain $D$
is said to be \emph{summable} if $f=g(n+1)-g(n)$ for some sequence $g\in D$. The problem of deciding whether
a given sequence is summable or not in~$D$ is called the \emph{summability problem} in~$D$.
For example, if $D$ is the field of rational functions, then for $f=1/(n(n+1))$ we can
find $g=1/n$, while for $f=1/n$ no suitable $g$ exists in~$D$. When $f$ is not summable
in~$D$, there are several other questions we may ask. One possibility is to ask
whether there is a pair~$(g, r)$ in~$D\times D$ such that $f(n) = g(n+1)-g(n) + r(n)$,
where $r$ is minimal in some sense and $r=0$ if~$f$ is summable.
This problem is called the \emph{decomposition problem} in~\cite{Abramov1995b}.

For univariate sequences, extensive work has been done to solve the summability and decomposition problems.
In 1971, Abramov solved the summability problem for univariate rational functions in~\cite{Abramov1971}.
The Gosper algorithm~\cite{Gosper1978} solves the summability problem for univariate hypergeometric terms.
This was then used by Zeilberger~\cite{Zeilberger1990c} in 1990s to design his celebrated telescoping algorithm for hypergeometric terms.
The Gosper algorithm was extended further to the $D$-finite case by
Abramov and van Hoeij in~\cite{AbramovHoeij1997, AbramovHoeij1999}, and to a more general difference-field setting
by Karr~\cite{Karr1981, Karr1985} and Schneider~\cite{Schneider2004}. The decomposition problem was first considered
by Ostrogradsky~\cite{Ostrogradsky1845} in 1845 and later by Hermite~\cite{Hermite1872} in the continuous setting for rational functions.
The discrete case was solved by Abramov in~\cite{Abramov1975}, with alternative methods
later presented by Abramov himself in~\cite{Abramov1995b}, and also by Paule~\cite{Paule1995b} and Pirastu~\cite{Pirastu1995a}. Abramov's decomposition algorithm was later extended to the
hypergeometric case in~\cite{Abramov2001, AbramovPetkovsek2002b}, as well as to continuous extensions in~\cite{BCCLX2013, CKK2016, Chen2017fuchsian}.

In 1993, Andrews and Paule~\cite{Andrews1993} raised the general question:
is it possible to provide any algorithmic device for reducing multiple sums to
single ones? This question is related to symbolic summation in the multivariate case.
To make the problem more tractable, we will focus on the first non-trivial case, namely
the bivariate rational functions. To this end, let us first introduce some notations.
Throughout the paper, let $k$ be a field of characteristic zero and $k(x, y)$ be the field of rational functions in~$x$ and~$y$.
For any~$f\in k(x, y)$, we define the shift operators~$\si_x, \si_y$ by
\[ \si_x(f(x, y)) = f(x+1, y), \quad \si_y(f(x, y)) = f(x, y+1), \]
and the $q$-shift operators with~$q\in k\setminus\{0\}$ by
\[ \tau_{x, q}(f(x, y)) = f(qx, y),\quad  \tau_{y, q}(f(x, y)) = f(x, qy).\]
Let~$\Delta_v:= \si_v-1$ and~$\Delta_{v, q}:= \tau_{v, q}-1$
be the difference and~$q$-difference operators with respect to~$v\in \{x, y\}$, respectively. On the field~$k(x, y)$, we can also define the usual derivations
$D_x := \partial/\partial_x$ and~$D _y:= \partial/\partial_y$.
\begin{definition}\label{DEF:exact}
A rational function~$f\in k(x, y)$ is said to be \emph{exact}
with respect to the pair~$(\partial_x, \partial_y ) \in \{D_x, \Delta_x, \Delta_{x, q}\}\times \{D_y, \Delta_y, \Delta_{y,q}\}$ in $k(x, y)$
if~$f = \partial_x(g) + \partial_y(h)$ for some~$g, h\in k(x, y)$.
\end{definition}

We study the following problem, which is a bivariate extension of the summability problem
for univariate rational functions.
\[
\begin{minipage}[t]{10.8cm}{\bf Exactness Testing Problem.}
    Given a rational function~$f\in k(x, y)$, decide whether or not~$f$ is exact with respect to $(\partial_x, \partial_y)$ in~$k(x, y)$.
\end{minipage}
\]

According to different types of $(\partial_x, \partial_y)$,  the above problem has six different cases up to the symmetry between $x$ and $y$.
In the pure continuous case, the problem is also called \emph{integrability problem},
which was first solved by Picard~\cite[vol 2, page 220]{Picard1897}, and see~\cite{ChenKauersSinger2012} for a more
up-to-date presentation. Chen and Singer~\cite{ChenSinger2014} presented the
first necessary and sufficient condition for the exactness in the pure discrete and $q$-discrete cases.
Based on the theoretical criterion in~\cite{ChenSinger2014}, Hou and Wang~\cite{HouWang2015} then
gave a practical algorithm for deciding the exactness in the corresponding case.
The goal of this paper is to solve the remaining three mixed cases of the exactness testing problem, which completes
our recent project on bivariate extensions of Abramov's algorithm for rational summation.

\section{Residues and reduced forms}  \label{SECT:red}

In this section, we will prepare some basic tools for testing the exactness of bivariate rational functions.
We first introduce the classical residues and their discrete analogue for univariate rational functions.
After this we will define reduced forms for bivariate rational functions.

Let $K$ be a field of characteristic zero and $K(z)$ be the field of rational functions
in $z$ over $K$. We first define residues with respect to the derivation $D_z$ on $K(z)$.
By irreducible partial fraction decomposition, we can always uniquely write a
rational function $f\in K(z)$ as
\begin{equation}\label{EQ:pfdc}
f = p+ \sum_{i=1}^n \sum_{j=1}^{m_i} \frac{a_{i, j}}{ d_i^j},
\end{equation}
where~$p, a_{i, j}, d_i \in K[z]$, $\deg_z(a_{i, j})< \deg_z(d_i)$ and all of the $d_i$'s
are distinct irreducible polynomials. We call $a_{i, 1}$ the \emph{$D_z$-residue} of
$f$ at $d_i$, denoted by $\res_{D_z}(f, d_i)$. We now recall the discrete analogue of
$D_z$-residues introduced in~\cite{ChenSinger2012, HouWang2015}.

Let $\phi$ be an automorphism of $K(z)$ that fixes $K$. For a polynomial $p\in K[z]$,
we call the set $\{\phi^i(p)\mid i\in \bZ\}$ the \emph{$\phi$-orbit} of $p$, denoted by $[p]_{\phi}$.
Two polynomials $p, q\in K[z]$ are said to be $\phi$-equivalent (denoted as $p\sim_{\phi} q$) if they are in the same $\phi$-orbit, i.e., $p = \phi^i(q)$ for some $i\in \bZ$.
When $\phi = \si_z$, we can uniquely decompose a rational function $f\in K(z)$ into the form
\begin{equation}\label{EQ:pfdd}
f = p(z) + \sum_{i=1}^n \sum_{j=1}^{m_i} \sum_{\ell=0}^{e_{i, j}} \frac{a_{i, j, \ell}}{ \si_z^\ell (d_i)^j},
\end{equation}
where $p, a_{i, j, \ell}, d_i\in K[z]$, $\deg_z(a_{i, j, \ell})< \deg_z(d_i)$ and all of the $d_i$'s
are irreducible polynomials such that any two of them are not $\si_z$-equivalent. We call the sum $\sum_{\ell=0}^{e_{i, j}} \si_z^{-\ell}(a_{i, j, \ell})$
the \emph{$\si_z$-residue} of $f$ at $d_i$ of multiplicity $j$, denoted by $\res_{\si_z}(f, d_i, j)$.

The following lemma shows some commutativity properties of the residues at some special irreducible polynomials.

\begin{lemma}\label{LEM:commute}
Let~$f = a/b \in k(x, y)$ and $d\in k[y]$ be an irreducible factor of $b$. Then the following commutativity formulae hold:
\begin{itemize}
\item[$(i)$]~ $\res_{D_y}(\si_x(f), d) = \si_x(\res_{D_y}(f, d))$;
\item[$(ii)$]~ $\res_{D_y}(\tau_{x, q}(f), d) = \tau_{x, q}(\res_{D_y}(f, d))$;
\item[$(iii)$]~~ $\res_{\si_y}(\tau_{x, q}(f), d, j) = \tau_{x, q}(\res_{\si_y}(f, d, j))$ for all $j\in \bN$.
\end{itemize}
\end{lemma}
\begin{proof}
To show the first formula, we decompose $f\in k(x, y)$ into the form
\[f = p + \sum_{i=1}^n \sum_{j=1}^{m_i} \frac{a_{i, j}}{ d_i^j}, \]
where $p, a_{i, j}\in k(x)[y],  d_i \in k[x, y]$ with $\deg_y(a_{i, j})< \deg_y(d_i)$ and the $d_i$'s
are distinct irreducible polynomials with $d_1 = d\in k[y]$. Since $\si_x$ is an automorphism of $k(x, y)$, we have that
\[\si_x(f) = \si_x(p) + \sum_{j=1}^{m_1} \frac{\si_x(a_{1, j})}{d_1^j} + \sum_{i=2}^n \sum_{j=1}^{m_i} \frac{\si_x(a_{i, j})}{ \si_x(d_i)^j}\]
is the irreducible partial fraction decomposition of $\si_x(f)$ with respect to $y$ over $k(x)$. Then $\res_{D_y}(\si_x(f), d) = \si_x(a_{1, 1}) = \si_x(\res_{D_y}(f, d))$.
The second formula can be proved similarly. To show the third formula, we decompose $f$ into the form
\[f = p + \sum_{i=1}^n \sum_{j=1}^{m_i} \sum_{\ell=0}^{e_{i, j}} \frac{a_{i, j, \ell}}{ \si_y^\ell (d_i)^j},\]
where $p, a_{i, j, \ell}\in k(x)[y], d_i\in k[x, y]$ with $\deg_y(a_{i, j, \ell})< \deg_y(d_i)$ and the $d_i$'s
are irreducible polynomials in distinct $\si_y$-orbits with $d_1 = d\in k[y]$. Since $\si_y$ is an automorphism of $k(x, y)$,
the polynomial $d\in k[y]$ is not $\si_y$-equivalent to any irreducible polynomial $d^\prime\in k[x, y]$
with $\deg_x(d^\prime)\neq 0$. Then we can decompose $\tau_{x, q}(f)$ into the form
\[\tau_{x, q}(f) = \tau_{x, q}(p) +  \sum_{j=1}^{m_1} \sum_{\ell=0}^{e_{1, j}} \frac{\tau_{x, q}(a_{1, j, \ell})}{ \si_y^\ell (d)^j} +  \frac{s}{t}, \]
where~$s\in k(x)[y]$ and $t\in k[x, y]$ satisfying that $\deg_y(s)<\deg_y(t)$ and any irreducible factor of $t$ is not $\si_y$-equivalent to $d$.
Then for all $j\in \bN$ we have
\[\res_{\si_y}(\tau_{x, q}(f), d, j) = \sum_{\ell=0}^{e_{1, j}} \si_y^{-\ell}{\tau_{x, q}(a_{1, j, \ell})}
= \tau_{x, q} \left(\sum_{\ell=0}^{e_{1, j}} \si_y^{-\ell}{(a_{1, j, \ell})}\right) = \tau_{x, q}(\res_{\si_y}(f, d, j)). \]
This completes the proof. \qed
\end{proof}

Let $\phi$ be any automorphism of $k(x, y)$ that fixes $k(y)$ which will be taken as $\tau_{x, q}$ or $\si_x$ in the next section.
Then $\phi$ commutes with $D_y$. To study the exactness testing problem with respect to the pair $(\phi, D_y)$,
we define reduced forms for rational functions in $k(x, y)$ as follows.

\begin{definition}
A rational function $r = \sum_{i=1}^{m} \frac{a_i}{d_i}$ with $a_i\in k(x)[y]$ and~$d_i\in k[x, y]$
is said to be \emph{$(\phi, D_y)$-reduced} if
$\deg_y(a_{i}) < \deg_y(d_i)$ and the $d_i$'s are irreducible polynomials in distinct $\phi$-orbits.
Let~$f\in k(x, y)$. We call the decomposition $f = \phi(g)-g + D_y(h) + r$ with $g, h, r\in k(x, y)$ and $r$ being $(\phi, D_y)$-reduced
a \emph{$(\phi, D_y)$-reduced form} of~$f$.
\end{definition}

We next show that $(\phi, D_y)$-reduced forms always exist for rational functions in $k(x, y)$.
For any rational function $f\in k(x, y)$, Ostrogradsky--Hermite reduction~\cite{Ostrogradsky1845, Hermite1872} decomposes $f$ into the form
\begin{equation}\label{EQ:df}
f = D_y(h) + \sum_{i=1}^m \frac{a_i}{d_i},
\end{equation}
where~$h\in k(x, y), a_i\in k(x)[y], d_i\in k[x, y]$ satisfying that $\deg_y(a_{i})<\deg_y(d_i)$ and
the $d_i$'s are irreducible over $k(x)$. Let~$\phi_1, \phi_2$ be two automorphisms of $k(x, y)$ such that
$\phi_1(\phi_2(f))=\phi_2(\phi_1(f))$ for all $f\in k(x, y)$.
Then for any~$a, d\in k(x)[y], m, n\in \bN$,  we have the following reduction formula
\begin{equation}\label{EQ:phi}
\frac{a}{\phi_1^m\phi_2^n (d)} = \phi_1(u) - u + \phi_2(v)-v+\frac{\phi_1^{-m}\phi_2^{-n}(a)}{d}
\end{equation}
where
\[ u = \sum_{j=0}^{m-1} \frac{\phi_1^{j-m}(a)}{\phi_1^j\phi_2^n(d)}\quad \text{and}
\quad v = \sum_{k=0}^{n-1}\frac{\phi_2^{k-n}\phi_1^{-m}(a)}{\phi_2^k(d)}.\]
By applying the above reduction formula to~\eqref{EQ:df} with $\phi_1=\phi$ and $\phi_2=id$, we can further decompose $f$ as
\[f = \phi(g)-g + D_y(h) + \sum_{i=1}^{\tilde m} \frac{\tilde a_i}{\tilde d_i},\]
where~$g\in k(x, y)$ and the $\tilde d_i$'s are in distinct $\phi$-orbits, which is a
$(\phi, D_y)$-reduced form of $f$. The above process for obtaining such a $(\phi, D_y)$-reduced form of $f$ is called a
\emph{$(\phi, D_y)$-reduction}.

Next we will define reduced forms for rational functions in $k(x, y)$ with respect to the pair~$(\tau_{x, q}, \si_y)$.
Two polynomials $p, p^\prime \in k[x, y]$ are said to be $(\tau_{x, q}, \si_y)$-equivalent if $p = \tau_{x, q}^m\si_y^n( p^\prime)$ for some $m, n \in \bZ$.
The set $\{\tau_{x, q}^{i}\si_y^j(p)\in k[x, y] \mid i, j \in \bZ\}$ is called the $(\tau_{x, q}, \si_y)$-orbit of $p$, denoted by $[p]_{(\tau_{x, q}, \si_y)}$.
\begin{definition}
A rational function $r = \sum_{i=1}^{n}\sum_{j=1}^{m_i} \frac{a_{i, j}}{d_i^j}\in k(x, y)$ with $a_{i, j}\in k(x)[y]$ and~$d_i\in k[x, y]$
is said to be \emph{$(\tau_{x, q}, \si_y)$-reduced} if
$\deg_y(a_{i, j}) < \deg_y(d_i)$ and the $d_i$'s are irreducible polynomials in distinct $(\tau_{x, q}, \si_y)$-orbits.
The decomposition $f = \Delta_{x, q}(g) + \Delta_y(h) + r$ with $g, h, r\in k(x, y)$ and $r$ being $(\tau_{x, q}, \si_y)$-reduced
is called a \emph{$(\tau_{x, q}, \si_y)$-reduced form} of~$f$.
\end{definition}

The existence of $(\tau_{x, q}, \si_y)$-reduced forms for rational functions relies on Abrramov's reduction~\cite{Abramov1995b} that
decomposes a rational function $f\in k(x, y)$ into the form
\[f = \Delta_y(h) + \sum_{i=1}^n  \sum_{j=1}^{m_i} \frac{a_{i, j}}{d_i^j}, \]
where~$h\in k(x, y), a_{i, j}\in k(x)[y], d_i\in k[x, y]$ satisfying that $\deg_y(a_{i, j})<\deg_y(d_i)$ and
the $d_i$'s are irreducible polynomials in distinct $\si_y$-orbits. Using
the formula~\eqref{EQ:phi} with $\phi_1 = \tau_{x ,q}$ and~$\phi_2=\si_y$, we can further decompose $f$ as
\[f = \Delta_{x, q}(g) + \Delta_y(h) + \sum_{i=1}^{\tilde n}  \sum_{j=1}^{\tilde m_i} \frac{a_{i, j}}{d_i^j},\]
where~$g\in k(x, y)$ and the $d_i$'s are in distinct $(\tau_{x, q}, \si_y)$-orbits, which is a
$(\tau_{x, q}, \si_y)$-reduced form of $f$. The above process for obtaining such a $(\tau_{x, q},\si_y)$-reduced form of $f$ is called a
\emph{$(\tau_{x, q}, \si_y)$-reduction}.


\section{Exactness criteria}  \label{SECT:criteria}

We first solve the exactness testing problem for the case in which $q\in k$ is a root of unity.
Assume that $m$ is the minimal positive integer such that~$q^m=1$ and $k$ contains all
$m$th roots of unity. For any $f\in k(x, y)$,
it is easy to show that~$\tau_{x, q}(f) = f$ if and only if~$f\in k(y)(x^m)$.
Note that $k(x, y)$ is a finite algebraic extension of $k(y)(x^m)$ of degree $m$.
We recall a lemma in~\cite{ChenSinger2014} on reduced forms for rational functions with respect to~$\tau_{x, q}$.
\begin{lemma}\label{LM:qru}
Let~$q$ be such that~$q^m=1$ with~$m$ minimal and let $f \in k(x, y)$.
\begin{enumerate}
\item[(a)] $f = \tau_{x ,q}(g) - g$ for some $g \in k(x, y)$ if and only if the trace $\Tr_{k(x, y)/k(y)(x^m)}(f) = 0$.
\item[(b)] Any rational function~$f\in k(x, y)$ can be decomposed into
\begin{equation}\label{rook}
f = \tau_{x, q}(g) - g + c, \quad \text{where~$g\in k(x, y)$ and~$c\in k(y)(x^m)$}.
\end{equation}
Moreover, $f$ is~$\tau_{x, q}$-summable in~$k(x, y)$ if and only if~$c=0$. We call this decomposition
a \emph{$\tau_{x, q}$-reduced form} for $f$.
\end{enumerate}
\end{lemma}

\begin{theorem}\label{THM:qru}
Let~$q$ be such that~$q^m=1$ with~$m$ minimal and let $f \in k(x, y)$. Assume that $f = \tau_{x, q}(g) - g + c$ with
$g\in k(x, y)$ and~$c\in k(y)(x^m)$ is a {$\tau_{x, q}$-reduced form} of $f$. Then $f$ is exact with respect to~$(\tau_{x, q}, \partial_y)$
with $\partial_y\in \{\Delta_y, D_y\}$ if and only if $c = \partial_y(d)$ for some $d\in k(y)(x^m)$.
\end{theorem}
\begin{proof}
The sufficiency is clear. To show the necessity, we assume that $f$ is exact with respect to~$(\tau_{x, q}, \partial_y)$
with $\partial_y\in \{\Delta_y, D_y\}$, so is $c$, i.e.,
$c = \Delta_{x, q}(u) + \partial_y(v)$ for some $u, v \in k(x, y)$.
Write $u = \sum_{i=0}^{m-1} u_i x^i$ and $v = \sum_{i=0}^{m-1}v_i x^i$ with $u_i, v_i \in k(y, x^m)$.
Then we have
\[c = u_1 (q-1)x + \cdots + u_{m-1} (q^{m-1}-1) x^{m-1} + \sum_{i=0}^{m-1} \partial_y(v_i) x^i. \]
Since $1, x, \ldots, x^{m-1}$ are linearly independent in $k(x, y)$ over $k(y, x^m)$, we get that
$c = \partial_y(v_0)$. \qed
\end{proof}

From now on, we assume that $q\in k\setminus \{0\}$ is not a root of unity.
For any $f\in k(x, y)$, we have $\tau_{x, q}(f) = f$ if and only if $f\in k(y)$.
We next solve the exactness testing problem in the case when $\partial_x\in \{\Delta_x, \Delta_{x, q}\}$ and $\partial_y = D_y$.

\begin{theorem}\label{THM:dqc}
Let~$\phi \in \{\si_x, \tau_{x, q}\}$ and $f \in k(x, y)$.
Assume that $f = \phi(g) - g + D_y(h) + \sum_{i=1}^m a_i/d_i$ with $a_i\in k(x)[y]$ and $d_i\in k[x, y]$
be a $(\phi, D_y)$-reduced form of $f$.
Then $f$ is exact with respect to~$(\partial_x, D_y)$ with $\partial_x = \phi -1$
if and only if for each~$i \in \{1, \ldots, m\}$, $d_i\in k[y]$ and $a_i = \partial_x(b_i)$ for some~$b_i\in k(x)[y]$.
\end{theorem}
\begin{proof}
The sufficiency is clear. To show the necessity, we assume that $f$ is exact with respect to~$(\partial_x, D_y)$.
This implies that $r = \sum_{i=1}^m a_i/d_i$ is also exact with respect to~$(\partial_x, D_y)$, i.e.,
$r = \phi(u) - u + D_y(v)$ for some $u, v\in k(x, y)$. By the Ostrogradsky--Hermite reduction, we first decompose $u$
into the form
\[u = D_y(\tilde u) + \sum_{i=1}^s \frac{v_i}{w_i},\]
where $\tilde u\in k(x, y)$, $v_i\in k(x)[y]$, and the $w_i$'s are irreducible polynomials in $k[x, y]$. Then we have
\[r = \sum_{i=1}^m \frac{a_i}{d_i} = T +
D_y(\tilde v) \quad \text{with $T = \sum_{i=1}^s \left(  \frac{\phi(v_i)}{\phi(w_i)} -  \frac{v_i}{w_i} \right)$ and~$\tilde v = \phi(\tilde u)- \tilde u + v .$}\]
Since $\phi$ is an automorphism of $k[x, y]$, the polynomials $\phi(w_i)$ are also irreducible and all of the simple fractions
in the irreducible partial fraction decomposition of $T$ have simple poles.

We first show that all of the $d_i$'s are in $k[y]$. Set $\cD := \{d_1, \ldots, d_m\}$ and $\cW := \{w_1, \ldots, w_s\}$.
Note that all of the simple fractions in $D_y(\tilde v)$ have at least double poles. This implies that $r = T$ and
each simple fraction $a_i/d_i$ can only be cancelled with some simple fractions of $T$.
Then for each $i \in \{1, ..., m\}$,  $d_i$ is equal to $w_{j_1}$ or $\phi(w_{j_1})$ for some $j_1\in \{1, \ldots, s\}$. Assume that $d_i = w_{j_1}$.
If $\phi(w_{j_1}) = w_{j_1}$, then $w_{j_1}\in k[y]$ by~\cite[Lemma 3.4]{ChenSinger2012}. Otherwise, $\phi(w_{j_1}) = w_{j_2}$
for some $j_2 \in \{1, \ldots, s\}\setminus \{j_1\}$. Indeed, If $\phi(w_{j_1}) = d_j$ with $i\neq j$, then $d_i$ is $\phi$-equivalent to $d_j$,
which contradicts with the assumption that the $d_i$'s are in distinct $\phi$-orbits. If $w_{j_2} = \phi(w_{j_2})$, we also get that $w_{j_2}$ is in~$k[y]$ and so is $d_i$.
Otherwise $\phi(w_{j_2}) = w_{j_3}$ for some $j_3 \in \{1, \ldots, s\}\setminus \{j_1, j_2\}$. Continuing this process, we either conclude that $d_i\in k[y]$ or
get a series of equalities
\[d_i = w_{j_1},  \phi(w_{j_1}) = w_{j_2}, \phi(w_{j_2}) = w_{j_3}, \ldots .  \]
Since the set $\cW$ is finite, there exists $t$ with $1\leq t \leq s$ such that $\phi(w_{j_t}) = w_{j_{\tilde t}}$ with $1\leq \tilde t \leq  t$.
Then $w_{j_{\tilde t}} = \phi^{t-\tilde t + 1}(w_{j_{\tilde t}})$, which implies that $w_{j_{\tilde t}}$ is in~$k[y]$ and so is $d_i$.
Similarly, we have $d_i\in k[y]$ when $d_i = \phi(w_{j_1})$.

Since $d_i\in k[y]$,  applying the commutativity formulae in Lemma~\ref{LEM:commute} yields
\begin{align*}
  a_i   =\res_{D_y}(r, d_i)   = \res_{D_y}(\phi(u)-u + D_y(v), d_i) = \res_{D_y}(\phi(u)-u, d_i)  = \phi(b_i) - b_i,
\end{align*}
where $b_i = \res_{D_y}(u, d_i) \in k(x)[y]$. \qed
\end{proof}

\begin{example} By Theorem~\ref{THM:dqc}, the rational function $1/(x+y)$ is not exact with respect to~$\Delta_x$ and~$D_y$ since $x+y$ is not in $k[y]$.
So is the rational function~$1/(xy)$ since $1/x \neq \Delta_x(g)$ for any $g\in k(x, y)$.
\end{example}

We now consider the exactness testing problem in the case when $\partial_x = \Delta_{x, q}$ and $\partial_y = \Delta_y$.
To this end, we first recall a lemma which is a special case of Lemma 5.4 in~\cite{Chen2015}.
\begin{lemma} \label{LM:intlin}
Let~$p$ be an irreducible polynomial in $k[x, y]$. Assume that $\tau_{x, q}^i\si_y^j(p)= p$ for some $i, j \in \bZ$ with $i\neq 0$.
Then $p\in k[y]$.
\end{lemma}

Let~$f\in k(x, y)$. We assume that $f = \Delta_{x, q}(g)+ \Delta_y(h) + r$ is a $(\tau_{x, q}, \si_y)$-reduced form of $f$.
Write $r= \sum_{i=1}^n\sum_{j=1}^{m_i} \frac{a_{i,j}}{d_i^j}$,
where $a_{i, j}\in k(x)[y]$ and $d_i\in k[x, y]$ satisfying that $\deg_y(a_{i, j})<\deg_y(d_i)$
and the $d_i$'s are in distinct $(\tau_{x, q}, \si_y)$-orbits.
Then $f$ is exact with respect to $(\Delta_{x, q}, \Delta_y)$ if and only if $r$ is exact with respect to $(\Delta_{x, q}, \Delta_y)$.
Note that the operators~$\tau_{x, q}$ and~$\si_y$ preserve the multiplicities of irreducible factors in the denominators of
rational functions.
Therefore the rational function~$r$ is exact with respect to $(\Delta_{x, q}, \Delta_y)$
if and only if for each~$j$, the rational function
\begin{equation}\label{EQ:rj}
r_j = \sum_{i=1}^m \frac{a_{i, j}}{d_i^j}
\end{equation}
is exact with respect to $(\Delta_{x, q}, \Delta_y)$. By the same argument in the proof of Lemma~3.2 in~\cite{HouWang2015},
$r_j$ is exact with respect to $(\Delta_{x, q}, \Delta_y)$ if and only if each simple fraction ${a_{i, j}}/{d_i^j}$
is exact with respect to $(\Delta_{x, q}, \Delta_y)$. We now give an exactness criterion
for rational functions of the form $a/d^m$.

\begin{lemma}\label{LM:qdd}
Let~$f =a/d^m$, where $m\in \bN$, $d\in k[x, y]$ is an irreducible polynomial and $a\in k(x)[y]$ is nonzero and $\deg_y(a)<\deg_y(d)$.
Then $f$ is exact with respect to $(\Delta_{x, q}, \Delta_y)$ if and only if $d\in k[y]$ and $a = \Delta_{x, q}(b)$ for some $b\in k(x)[y]$.
\end{lemma}

\begin{proof} The sufficiency is clear.
For the necessity, we will outline the same argument used in the proof of Theorem 3.7 in~\cite{ChenSinger2014} or that of Proposition 3.4 in~\cite{HouWang2015}.
We assume that~$f$ is exact with respect to $(\Delta_{x, q}, \Delta_y)$, i.e., there exist~$g, h\in k(x, y)$ such that
\begin{equation}\label{EQ:thm}
f = \Delta_{x, q}(g)+ \Delta_y(h).
\end{equation}
We decompose the rational function~$g$ into the form
\begin{equation}\label{EQ:thm2}
g = \si_y(g_1)-g_1 + g_2 + \frac{\lambda_1}{\tau_{x, q}^{\mu_1} d^m} + \cdots +\frac{\lambda_s}{\tau_{x, q}^{\mu_s} d^m},
\end{equation}
where~$\lambda_k\in k(x)[y], \mu_k \in \bZ$, $g_1, g_2\in k(x, y)$ such that~$g_2$ is a rational function
having no terms of the form $\lambda/(\tau_{x, q}^{\mu} d^m)$ in its partial fraction decomposition with respect
to $y$, and the~$(\tau_{x, q}^{\mu_i} d^m)$'s are irreducible polynomials in distinct~$\si_y$-orbits.

The following claim can be shown by the same argument as in~\cite{ChenSinger2014, HouWang2015}.

\medskip
\noindent {\bf Claim 1.} Let
\[\Lambda := \{\tau_{x, q}^{\mu_1} d, \ldots, \tau_{x, q}^{\mu_s} d, \tau_{x, q}^{\mu_1+1} d, \ldots, \tau_{x, q}^{\mu_s+1} d\}.\]
Then: (1) at least one element of $\Lambda$ is in the same $\si_y$-orbit as $d$; (2) for each~$\eta \in \Lambda$, there is one
element of $(\Lambda\backslash \{\eta\}) \cup \{ d\}$ that is $\si_y$-equivalent to~$\eta$.

\smallskip
Claim 1 implies that either $d \sim_{\si_y} \tau_{x, q}^{\mu'_1} d$ or $d \sim_{\si_y} \tau_{x, q}^{\mu'_1+1} d$ for
some $\mu'_1 \in \{\mu_1, \ldots, \mu_s\}$. Assume that $d \sim_{\si_y} \tau_{x, q}^{\mu'_1} d$.  By the same argument as in~\cite{ChenSinger2014, HouWang2015}, we can show that
there exists a positive integer $t\leq s$ and $j\in \bZ$ such that $\tau_{x, q}^t \si_y^j(d)=d$, which implies $d\in k[y]$ by Lemma~\ref{LM:intlin}.
Similarly, if~$d \sim_{\si_y} \tau_{x, q}^{\mu'_1+1} d$, then we also have $d\in k[y]$.

Since $d\in k[y]$,  applying the commutativity formulae in Lemma~\ref{LEM:commute} yields
\begin{align*}
  a   =\res_{\si_y}(f, d, m)   = \res_{\si_y}( \Delta_{x, q}(g) + \Delta_y(h), d, m) = \res_{\si_y}(\Delta_{x, q}(g), d, m)  = \Delta_{x, q}(b),
\end{align*}
where $b = \res_{\si_y}(g, d, m) \in k(x)[y]$. \qed
\end{proof}

We conclude the above discussions by the following theorem.

\begin{theorem}\label{THM:dqd}
Let $f \in k(x, y)$ and assume that
\[f = \Delta_{x, q}(g)+ \Delta_y(h) + \sum_{i=1}^n\sum_{j=1}^{m_i} \frac{a_{i,j}}{d_i^j}\]
with $a_{i, j}\in k(x)[y]$ and $d_i\in k[x, y]$ is a $(\tau_{x, q}, \si_y)$-reduced form of $f$.
Then $f$ is exact with respect to the pair~$(\Delta_{x, q}, \Delta_y)$
if and only if for each~$i \in \{1, \ldots, n\}$, $d_i\in k[y]$ and  for each $j\in \{1, \ldots, m_i\}$,
$a_{i, j} = \Delta_{x, q}(b_{i, j})$ for some~$b_{i, j}\in k(x)[y]$.
\end{theorem}

\begin{example} By Theorem~\ref{THM:dqd}, the rational function $1/(x+y)$ is not exact with respect to~$\Delta_{x, q}$ and~$\Delta_y$ since $x+y$ is not in $k[y]$.
But the rational function~$1/(xy)$ is exact with respect to~$\Delta_{x, q}$ and~$\Delta_y$. In fact, $\frac{1}{xy} = \Delta_{x, q}\left(\frac{q}{(1-q)xy}\right)$.
\end{example}

\begin{remark}
The exactness criteria given above reduce the exactness testing problem in the bivariate case to two subproblems: one is testing whether an irreducible polynomial $p\in k[x, y]$
is free of $x$, the other is testing whether a rational function is $(q)$-summable or not with respect to $x$. The first subproblem is easy and the second one
 can be solved by Abramov's algorithm and its $q$-analogue for univariate rational summation.
\end{remark}

\section{Conclusion} \label{SECT:conc}
We conclude this paper by recalling the following open problem proposed in~\cite{ChenKauers2017}:
\begin{problem}
   Develop an algorithm which takes as input a multivariate hypergeometric term
   $h$ in $m$ discrete variables $k_1,\dots,k_m$, and decides whether there exist
   hypergeometric terms $g_1,\dots,g_m$ such that
   \[
     h = \Delta_1(g_1) + \cdots + \Delta_m(g_m).
   \]
   Here, $\Delta_i$ is the forward difference operator with respect to
   the variable~$k_i$, i.e., \[\Delta_i
   f(k_1,\dots,k_m)=f(k_1,\dots,k_i+1,\dots,k_m)-f(k_1,\dots,k_i,\dots,k_m).\]
 \end{problem}
A solution of this problem would be an important step towards the development
of a Zeilberger-like algorithm for multisums. Together with the results in~\cite{ChenSinger2014, HouWang2015},
the exactness criteria in previous section enable us to completely solve the above problem in
the case of bivariate rational functions. The summability criteria in~\cite{ChenSinger2014, HouWang2015}
were used in~\cite{ChenHLW2016} to derive some conditions on the existence of telescopers for trivariate rational functions.
Hopefully, the results in this paper can be used to solve the
corresponding existence problems for the three mixed cases.
An answer to the above open problem  may
analogously allow for the formulation of existence criteria for telescopers in
the multivariate setting. In the long run, we would hope that a multivariate Gosper algorithm
serves as a starting point for the development of a reduction-based creative telescoping algorithm
for the multivariate setting. A necessary condition for bivariate
hypergeometric summability has been given in~\cite{ChenHouMu2006} with many applications but the summability criterion in this
case is still missing and further new ideas and tools are needed to be developed.

\bigskip
\noindent {\bf Acknowledgment.}~
I would like to thank Hui Huang and Rong-Hua Wang for their constructive comments on the early version of this paper.

\bibliographystyle{plain}
\def\cprime{$'$}

\end{document}